\documentclass[aps,prc,preprint,amsmath,amssymb,showpacs,preprintnumbers,superscriptaddress]{revtex4-1}

\usepackage[utf8]{inputenc}
\usepackage[sort&compress]{natbib}
\usepackage{ulem}
\usepackage{bm}
\usepackage{times}
\usepackage{amssymb,amsbsy,amsmath,amsfonts}
\usepackage{graphicx}
\usepackage{float}
\usepackage{color}
\usepackage{morefloats}
\usepackage{rotating}
\usepackage{srcltx}
\usepackage{slashed}
\usepackage{subfigure}
\usepackage{multirow}
\usepackage{verbatim}
\usepackage{hyperref}
\usepackage{tabularx}
\usepackage{braket}

\DeclareUnicodeCharacter{3000}{HEREHEREHERE}

\begin{document}

\title{Nuclear and neutron matter in the relativistic Brueckner-Hartree-Fock theory with next-to-leading order covariant chiral nuclear force}% Force line breaks with \\

\author{Wei-Jiang Zou}
\affiliation{State Key Laboratory of Nuclear Physics and Technology, School of Physics, Peking University, Beijing 100871, China}
\author{Yi-Long Yang}
\affiliation{State Key Laboratory of Nuclear Physics and Technology, School of Physics, Peking University, Beijing 100871, China}
\author{Jun-Xu Lu}
\affiliation{School of Physics, Beihang University, Beijing 102206, China}
\author{Peng-Wei Zhao}
\affiliation{State Key Laboratory of Nuclear Physics and Technology, School of Physics, Peking University, Beijing 100871, China}
\author{Li-Sheng Geng}
\email{lisheng.geng@buaa.edu.cn}
\affiliation{School of Physics, Beihang University, Beijing 102206, China}
\affiliation{Sino-French Carbon Neutrality Research Center, \'Ecole Centrale de P\'ekin/School of General Engineering, Beihang University, Beijing 100191, China}
\affiliation{Peng Huanwu Collaborative Center for Research and Education, Beihang University, Beijing 100191, China}
\affiliation{Beijing Key Laboratory of Advanced Nuclear Materials and Physics, Beihang University, Beijing 102206, China }
\affiliation{Southern Center for Nuclear-Science Theory (SCNT), Institute of Modern Physics, Chinese Academy of Sciences, Huizhou 516000, China}

\author{Jie Meng}
 \email{mengj@pku.edu.cn}
\affiliation{State Key Laboratory of Nuclear Physics and Technology, School of Physics, Peking University, Beijing 100871, China}
\affiliation{Center for Theoretical Physics, China Institute of Atomic Energy, Beijing, 102413, China}

\date{\today}

\begin{abstract}

The symmetric nuclear matter and pure neutron matter are investigated by the relativistic Brueckner-Hartree-Fock (RBHF) theory 
with the covariant chiral nuclear forces up to the next-to-leading order~(NLO).
A fitting scheme to ensure the naturalness of the low-energy constants is proposed, 
which plays a crucial role in the proper description of nuclear matter.
With a momentum cutoff $\Lambda=590$ MeV, the empirical saturation energy and density, 
as well as the incompressibility coefficient at the saturation density are reproduced well.
The EoSs show less dependence on the momentum cutoff and become softer at densities above saturation density, 
in comparison with the previous leading order results.
Given the good description for the saturation properties of nuclear matter, the present work encourages future studies of 
the finite nuclei in the framework of the RBHF theory with the NLO covariant chiral nuclear forces.

\end{abstract}

\maketitle

\section{Introduction}

Nuclear matter is an essentially infinite nucleus with given ratio of neutrons to protons, 
and with the Coulomb force between the protons omitted.
The saturation properties of nuclear matter and its EoS around the saturation density provide useful benchmarks
for nuclear force models and many-body theories \cite{meng2016relativistic,Roca-Maza:2018ujj,Shen:2019dls,Tong:2019juo}.
A reliable microscopic theory for finite nuclei requires the underlying nuclear force to correctly describe nuclear matter.

However, it remains a nontrivial task to properly describe the saturation properties of nuclear matter 
in nuclear \textit{ab initio} approaches. 
In such an approach, nuclear matter is solved through nuclear many-body methods starting from 
realistic nucleon-nucleon ($NN$) interactions fixed by free-space scattering data.
It was found rather early that all the non-relativistic $NN$ interactions, though achieving high precision 
in describing free-space scattering data, failed to reproduce the saturation properties of nuclear matter.
The saturation binding energies and the corresponding densities obtained with various non-relativistic $NN$ interactions align along 
the so-called Coester line \cite{PhysRevC.1.769}, which significantly deviates from the empirical value.
Therefore, one needs to supplement non-relativistic $NN$ interactions with three-nucleon forces (3NF) to reasonably 
describe the saturation properties of nuclear matter.
Nevertheless, the predicted saturation properties of nuclear matter depend sensitively on 
the model of 3NF \cite{Coraggio:2014nva,Sammarruca:2014zia,Lonardoni:2019ypg} (for a recent review, see Ref.\cite{Hebeler:2020ocj})~and 
their proper description often relies on manual adjustments of 3NF \cite{Drischler2019PRL}.
In addition, a consistent 3NF for both finite nuclei and nuclear matter remains a challenge \cite{Sammarruca2020PRC, Huther2020PLB}.
More recently, many \textit{ab initio} calculations of nuclear structure employing chiral two- and three-nucleon interactions found 
that the properties of medium-mass nuclei and the saturation properties of nuclear matter 
cannot be consistently described \cite{Hoppe:2019uyw,Huther:2019ont,Machleidt:2023jws}.
This has led some groups to fit nuclear forces not only to the data of free-space scattering or light nuclei 
but also to the properties of medium-mass nuclei \cite{Ekstrom:2015rta}, or directly to the saturation properties of nuclear matter \cite{Drischler:2017wtt}. 
One such example is the N${}^2$LO${}_{\mathrm{SAT}}$ \cite{Ekstrom:2015rta} in which the ground-state energies and 
radii of oxygen isotopes are simultaneously included in the fitting to obtain an \textit{NN}+3\textit{N} interaction at the next-to-next-to-leading order (N$^2$LO).

An alternative approach is to employ relativistic \textit{ab initio} methods. 
In Refs.\cite{Brockmann:1990cn,Wang:2021mvg}, with the Bonn $NN$ potentials \cite{Machleidt:1989tm} only, 
the relativistic Brueckner-Hartree-Fock (RBHF) theory can yield the saturation properties of nuclear matter 
that are closer to the empirical value, compared to the non-relativistic Brueckner-Hartree-Fock (BHF) theory.
Furthermore, the recent relativistic quantum Monte Carlo calculations of light nuclei \cite{Yang2025ChinesePhysicsLetters051201} 
revealed that the light nuclei and the saturation properties of nuclear matter can be simultaneously described with the Bonn potentials.
Apart from these achievements, substantial progress has been made in relativistic \textit{ab initio} approaches \cite{Ring2023JPCS, Qin2024PRC, Yang2022PLB, Shen2018PRC, Shen2016CPL} in recent years.

However, the above-mentioned Bonn potentials are based on the one-boson exchange model \cite{Machleidt:1989tm}, 
which has no clear relation to the Quantum Chromodynamics (QCD), the underlying theory of the strong interaction.
In addition, they do not provide a systematic way to improve the description of nuclear force.
From this perspective, the chiral effective field theory ($\chi$EFT) can provide QCD-based $NN$ interactions 
for relativistic \textit{ab initio} calculations, since $\chi$EFT is based on the most general effective Lagrangian 
consistent with the chiral symmetry of QCD and its spontaneous and explicit breakings.
In the past, most $\chi$EFT studies of $NN$ interactions focused on non-relativistic $NN$ interactions \cite{Epelbaum:2008ga,Machleidt:2011zz}. 
Recent studies have proposed that chiral $NN$ interactions derived from the covariant $\chi$EFT~\cite{Ren:2016jna,Lu:2021gsb} can 
provide inputs for relativistic \textit{ab initio} calculations.
In Ref.\cite{Zou:2023quo}, it was demonstrated that the RBHF theory with the leading order (LO) covariant chiral $NN$ interactions 
already have the capacity to reasonably describe the saturation properties of nuclear matter.
This is in sharp contrast to the non-relativistic case where the saturation properties of nuclear matter have to 
be achieved with the additional 3NFs at the N$^2$LO.
The single-particle potential $U_\Lambda$ for $\Lambda$ hyperon at the saturation density, calculated using the RBHF theory 
with LO covariant chiral hyperon-nucleon ($YN$) and $NN$ interactions \cite{Zheng:2025sol}, aligns well with the empirical value.

In this work, we study the EoSs of symmetric nuclear matter (SNM) and pure neutron matter (PNM) in the RBHF theory 
with covariant chiral $NN$ interactions up to the next-to-leading order (NLO). 
To determine the low-energy constants (LECs) in the covariant chiral nuclear forces, we develop a fitting procedure to the $NN$ scattering phase shifts 
that ensures the naturalness of the LECs, which is crucial for a proper description of the nuclear matter EoSs in the present RBHF theory. 
The cutoff dependence of the EoSs is studied by varying the momentum cutoff scale in the range $\Lambda=450\text{-}600$ MeV. 
This paper is organized as follows. In Sec.\ref{sec:level2}, we briefly introduce the theoretical framework of 
the $NN$ interactions in the covariant chiral EFT up to NLO and the RBHF theory for nuclear matter. 
In Sec.\ref{sec:level3}, we present the fitting procedure of the LECs in the covariant chiral nuclear forces 
and the RBHF results for the EoSs of SNM and PNM. 
Finally, we provide a summary and outlook in Sec.\ref{sec:level4}.

\section{Theoretical framework}\label{sec:level2}
\subsection{Covariant chiral $NN$ interactions}
We start from the $NN$ interactions derived within the covariant chiral EFT.
Here and in what follows, $q \in \{m_\pi/\Lambda_b, \; |\bm p \, |/\Lambda_b\}$ denotes the chiral expansion parameter, 
with $m_\pi$, $\vec p$ and $\Lambda_b$ referring to the pion mass, a typical nucleon momentum, and the breakdown scale, respectively.
Up to NLO, the covariant chiral potential~$V$~consists of the following terms \cite{Lu:2021gsb},
\begin{align}
  V =   V_{\mathrm{CT}}^{\mathrm{LO}} 
      + V_{\mathrm{CT}}^{\mathrm{NLO}}
      + V_{\mathrm{OPE}}
      + V_{\mathrm{TPE}}^{\mathrm{NLO}}
      - V_{\mathrm{IOPE}},
  \end{align}
in which the first two terms refer to the LO [$\mathcal{O}(q^0)$] and NLO [$\mathcal{O}(q^2)$] contact contributions, 
while the next three terms denote one-pion-exchange (OPE) contribution, leading two-pion-exchange (TPE) contribution 
and the iterated OPE contribution, respectively. 
The pion exchange potentials have been determined in Ref.\cite{Wang:2021kos}.
\begin{table*}[htbp]
\centering
\caption{A complete set of $NN$ contact interaction terms up to $\mathcal{O}(q^2)$ in the covariant chiral EFT.}\label{Table1}
\scalebox{0.82}{
\begin{tabular}{cccc}
\hline\hline
${\tilde{O}_1}$  &   $\bar{u}(\bm{p}')u(\bm{p})\bar{u}(-\bm{p}')u(-\bm{p})$                                                                                                    &         $\tilde{O}_{10}$       &      $\frac{-1}{4M^2}(\bar{p}'-\bar{p})^2\bar{u}(\bm{p}')u(\bm{p})\bar{u}(-\bm{p}')u(-\bm{p})$\\
${\tilde{O}_2}$  &   $\bar{u}(\bm{p}')\gamma^\mu u(\bm{p})\bar{u}(-\bm{p}')\gamma_\mu u(-\bm{p})$                                                                              &         $\tilde{O}_{11}$       &      $\frac{-1}{4M^2}(\bar{p}'-\bar{p})^2\bar{u}(\bm{p}')\gamma^\mu u(\bm{p})\bar{u}(-\bm{p}')\gamma_\mu u(-\bm{p})$\\
${\tilde{O}_3}$  &   $\bar{u}(\bm{p}')\gamma_5\gamma^\mu u(\bm{p})\bar{u}(-\bm{p}')\gamma_5\gamma_\mu u(-\bm{p})$                                                              &          $\tilde{O}_{12}$      &    
$\frac{-1}{4M^2}(\bar{p}'-\bar{p})^2\bar{u}(\bm{p}')\gamma_5\gamma^\mu u(\bm{p})\bar{u}(-\bm{p}')\gamma_5\gamma_\mu u(-\bm{p})$\\
${\tilde{O}_4}$  &   $\bar{u}(\bm{p}')\sigma^{\mu\nu}u(\bm{p})\bar{u}(-\bm{p}')\sigma_{\mu\nu}u(-\bm{p})$                                                                      &          $\tilde{O}_{13}$      &    $\frac{-1}{4M^2}(\bar{p}'-\bar{p})^2\bar{u}(\bm{p}')\sigma^{\mu\nu}u(\bm{p})\bar{u}(-\bm{p}')\sigma_{\mu\nu}u(-\bm{p})$\\
$\tilde{O}_5$    &   $\bar{u}(\bm{p}')\gamma_5u(\bm{p})\bar{u}(-\bm{p}')\gamma_5u(-\bm{p})$                                                                                    &          $\tilde{O}_{14}$      &     $\frac{1}{4M^2}\bar{u}(\bm{p}')(p'+p)^\alpha u(\bm{p})\bar{u}(-\bm{p}')(\bar{p}'+\bar{p})_\alpha u(-\bm{p})-\tilde{O}_1$\\
$\tilde{O}_6$    &   $\frac{1}{4M^2}\bar{u}(\bm{p}')\gamma_5\gamma^\mu (p'+p)^\alpha u(\bm{p})\bar{u}(-\bm{p}')\gamma_5\gamma_\alpha(\bar{p}'+\bar{p})_\mu u(-\bm{p})$         &          $\tilde{O}_{15}$      &    
$\frac{1}{4M^2}\bar{u}(\bm{p}')\gamma^\mu(p'+p)^\alpha u(\bm{p})\bar{u}(-\bm{p}')\gamma_\mu(\bar{p}'+\bar{p})_\alpha u(-\bm{p})-\tilde{O}_2$\\
$\tilde{O}_7$    &   $\frac{1}{4M^2}\bar{u}(\bm{p}')\sigma^{\mu\nu}(p'+p)^\alpha u(\bm{p})\bar{u}(-\bm{p}')\sigma_{\mu\alpha}(\bar{p}'+\bar{p})_\nu u(-\bm{p})$                &          $\tilde{O}_{16}$      &    $\frac{1}{4M^2}\bar{u}(\bm{p}')\gamma_5\gamma^\mu(p'+p)^\alpha u(\bm{p})\bar{u}(-\bm{p}')\gamma_5\gamma_\mu(\bar{p}'+\bar{p})_\alpha u(-\bm{p})-\tilde{O}_3$\\
$\tilde{O}_8$    &   $\frac{i}{4M^2}\bar{u}(\bm{p}')(p'+p)^\mu u(\bm{p})\bar{u}\sigma_{\mu\nu}(\bar{p}'-\bar{p})^\nu u(-\bm{p})$                                               &          $\tilde{O}_{17}$      &    
$\frac{1}{4M^2}\bar{u}(\bm{p}')\sigma^{\mu\nu}(p'+p)^\alpha u(\bm{p})\bar{u}(-\bm{p}')\sigma_{\mu\nu}(\bar{p}'+\bar{p})_\alpha u(-\bm{p})-\tilde{O}_4$\\
$\tilde{O}_9$    &   $\frac{-1}{4M^2}\bar{u}(\bm{p}')\sigma^{\mu\alpha}u(\bm{p})\bar{u}(-\bm{p}')\sigma_{\mu\nu}(\bar{p}'-\bar{p})_\alpha(\bar{p}'-\bar{p})^\nu u(-\bm{p})$    &          $\qquad$              &   
$\qquad$\\
\hline\hline
\end{tabular}}
\end{table*}

The contact interactions take the form \cite{Xiao:2018jot}
\begin{equation}
    V_{\rm CT}^{\rm LO}  = \sum_{i=1}^4    \tilde{C}_i\tilde{O}_i,\quad 
    V_{\rm CT}^{\rm NLO} = \sum_{i=5}^{17} \tilde{C}_i\tilde{O}_i.
\end{equation}
There are, in total, 4 contact terms at LO and 17 contact terms up to NLO.
For each contact term $\tilde{O}_i$, there is an associated LEC $\tilde{C}_i$ to be 
determined by fitting to the $NN$ observables.
The momentum-space matrix elements of the contact terms are shown in Table \ref{Table1}. 
The initial four-momenta $p$ and $\bar{p}$ in the matrix element expression are defined as
\begin{align}
      p^\mu = (E_{p}, \bm{p}),\qquad 
\bar{p}^\mu = (E_p  ,-\bm{p}).
\end{align}
$p'$ and $\bar{p}'$ are the corresponding final four-momenta. 
Since the initial and final momenta are taken to be on the mass shell.
The two-body interaction matrix elements can all be expressed solely as functions of the three-momenta.
Therefore, this form of the interaction does not involve energy dependence, making it suitable for many-body calculations.

To avoid ultraviolet divergences, the covariant chiral potential $V$ is regularized by the following separable Gaussian-like function
\begin{equation}
    f_\Lambda(p',p)=\exp\left(\frac{-\bm{p}^{2n}-\bm{p}^{\prime2n}}{\Lambda^{2n}}\right),
\end{equation}
with $n=3$ and $\Lambda$ is the momentum cutoff.

However, it should be noted that the LECs determined in Ref.\cite{Lu:2021gsb} are not fully consistent with the framework 
adopted in the present work. 
Specifically, whereas Ref.\cite{Lu:2021gsb} employs the Blankenbecler-Sugar (BbS) equation and sharp form factor, 
we solve the Thompson equation to take into account the non-perturbation effects, as shown in the next subsection, 
and adopt Gaussian-like form factors to remove divergences. 
Besides, the interactions in this work are simplified in an energy-independent way. 
Another point, which may be more important, is that solely fitting to $NN$ phase shifts without any other constraints, 
e.g., the requirements of naturalness of the LECs, lead to instabilities in the RBHF calculations. 
Considering all the above points, we need to refit the LECs under the constraints of naturalness, similar to Ref.\cite{Epelbaum:2014efa}.

On the other hand, the natural size for LECs is closely related to the power counting rules adopted to organize the chiral nuclear force. 
In the non-relativistic framework \cite{Epelbaum:2014efa}, the natural size for LECs in various orders is estimated 
with the order of the pertinent hard scale $\Lambda_b$ following Weinberg's power counting, which is based on the naive dimensional analysis. 
In the present situation, however, the covariant power counting rules dictate that in addition to the non-relativistic operators of the corresponding order, 
it also contains operators of higher orders as per the non-relativistic power counting rules. 
The LECs in the covariant chiral nuclear force are a combination of contributions from various orders of hard scale $\Lambda_b$. 
Consequently, the estimation of their natural size is non-trivial. 
Nevertheless, we simply follow the estimation of non-relativistic chiral nuclear force in this work.

\subsection{The RBHF theory}

In the RBHF theory, the basic quantity is the $G$ matrix, which is the effective interaction between the nucleons.
The $G$ matrix is the solution of the in-medium scattering equation and is employed in relativistic Hartree-Fock calculations.
One of the most widely used scattering equations in the RBHF theory is the Thompson equation \cite{Thompson:1970wt}, 
namely a relativistic three-dimensional reduction of the Bethe-Salpeter equation \cite{Salpeter:1951sz},
\begin{eqnarray}
    G(\bm{q}',\bm{q}|\bm{P},W)
    =&&        V(\bm{q}',\bm{q}|\bm{P})
         +    \int\frac{\mathrm{d}^3k}{(2\pi)^3}V(\bm{q}',\bm{k}|\bm{P})\nonumber\\
    &&\times  \frac{M^{\ast2}}{E_{\bm{P}+\bm{k}}^\ast E_{\bm{P}-\bm{k}}^\ast}
              \frac{Q(\bm{k},\bm{P})}{W-E_{\bm{P}+\bm{k}}-E_{\bm{P}-\bm{k}}}\nonumber\\
    &&\times  G(\bm{k},\bm{q}|\bm{P},W)\label{equ9},
\end{eqnarray}
where $\bm{P}=\frac{1}{2}(\bm{k}_1+\bm{k}_2)$ is the center-of-mass momentum, and $\bm{k}=\frac{1}{2}(\bm{k}_1-\bm{k}_2)$ is the relative momentum of 
the two interacting nucleons with momenta $\bm{k}_1$ and $\bm{k}_2$, and $\bm{q}$, $\bm{q}'$, and $\bm{k}$ are the initial, final, and intermediate 
relative momenta of the two interacting nucleons in nuclear matter, respectively. 
The starting energy $W=E_{\bm{P}+\bm{q}}+E_{\bm{P}-\bm{q}}$, 
$M^\ast$~and~$E^\ast_{\bm{P}\pm\bm{k}}$~are effective mass and energies which will be specified below.
The Pauli operator $Q(\bm{k},\bm{P})$ only allows the scattering of nucleons to unoccupied states, i.e.,
\begin{eqnarray}
    Q(\bm{k},\bm{P})
    =\begin{cases}
        1,\quad      |\bm{P}+\bm{k}|,|\bm{P}-\bm{k}|>k_F\\
        0,\quad      \text{otherwise}
    \end{cases},
\end{eqnarray}
where $k_F$ is the Fermi momentum.

In the RBHF theory, the single-particle motion of the nucleon in nuclear matter is described by the Dirac equation
\begin{eqnarray}\label{eq.dirac}
(\bm{\alpha}\cdot\bm{p} +  \beta M+  \beta\mathcal{U})  u(\bm{p},\lambda)
                                            =E_{\bm{p}} u(\bm{p},\lambda),
\end{eqnarray}
where $\bm{\alpha}$ and $\beta$ are the Dirac matrices, $u(\bm{p},\lambda)$ is the Dirac spinor with momentum $\bm{p}$, 
single-particle energy $E_{\bm{p}}$ and helicity $\lambda$, $M$ is the mass of the free nucleon, 
and~$\mathcal{U}$~is the single-particle potential operator, providing the primary medium effects. 
Due to the time-reversal invariance, one can neglect the spacelike component of the vector fields, and as a result, 
the single-particle potential operator can be expressed as \cite{Brockmann:1990cn,Tong:2018qwx}:   
\begin{eqnarray}
    \mathcal{U} = U_S + \gamma^0U_0,
\end{eqnarray}
The momentum dependence of the scalar field ($U_S$) and the timelike component of 
the vector field ($U_0$) is neglected \cite{Brockmann:1990cn,Tong:2018qwx}.
The solution of the in-medium Dirac equation can be expressed in the form of the Dirac spinor with effective quantities,
\begin{eqnarray}
    u(\bm{p},\lambda) = \sqrt{\frac{E^\ast_{\bm{p}}+M^\ast}{2M^\ast}}
                        \begin{pmatrix}                                 
                                             1                       \\ 
                         \dfrac{2\lambda p}{E^\ast_{\bm{p}}+M^\ast} 
                        \end{pmatrix}  
                         \chi_{\lambda},
\end{eqnarray}
where $\chi_{\lambda}$ is the Pauli spinor in helicity basis and 
\begin{eqnarray}
    M^\ast          = M          + U_S,\quad 
    E^\ast_{\bm{p}} = E_{\bm{p}} - U_0.
\end{eqnarray}
The covariant normalization $\bar{u}(\bm{p},\lambda)u(\bm{p},\lambda)=1$ is adopted.

The single-particle scalar and vector potentials $U_S$ and $U_0$ are determined from the $G$ matrix.
One first obtain the single-particle potential operator
\begin{widetext}
\begin{eqnarray}\label{eq.sigma}
    \Sigma(p)   = \sum_{\lambda'}
                  \int_0^{k_F}\frac{\mathrm{d}^3p'}{(2\pi)^3}\frac{M^\ast}{E_{\bm{p}'}^\ast}
                  \langle \bar{u}(\bm{p},1/2)
                  \bar{u}(\bm{p}',\lambda')
                 |\bar{G}|u(\bm{p},1/2)u(\bm{p}',\lambda')\rangle\label{equ8},
\end{eqnarray}
\end{widetext}
where $\bar{G}$ is the antisymmetrized $G$ matrix. 
Within the no-sea approximation \cite{Serot:1997xg}, the integral is only performed for the single-particle states in the Fermi sea.

The potentials $U_S$~and~$U_0$ are determined by the following relations \cite{Anas83,Brockmann:1990cn} ,
\begin{subequations}\label{equ7}
\begin{eqnarray}
  \Sigma(p_1) = U_S + \frac{E_{p_1}^\ast}{M^\ast}U_0,\\
  \Sigma(p_2) = U_S + \frac{E_{p_2}^\ast}{M^\ast}U_0,
\end{eqnarray}
\end{subequations}
where $p_1 = 0.5k_F$ and $p_2=0.7k_F$ are two momenta in the Fermi sea.
Using the Bonn potentials, the dependence on the momenta $p_1$ and $p_2$ is investigated for the EoS of the SNM \cite{Wang:2021mvg}.
For the covariant chiral nuclear forces in this paper, the momentum independence approximation is adopted.

The coupled equations.\eqref{equ9}, \eqref{eq.dirac}, \eqref{eq.sigma}, \eqref{equ7} can be solved in a self-consistent way.
Starting from the initial $U_S^{(0)}$ and $U_0^{(0)}$, the Dirac equation \eqref{eq.dirac} is solved to obtain the Dirac spinors. 
With the Dirac spinors, the $G$ matrix is obtained by the Thompson equation \eqref{equ9}, and accordingly the self-energy $\Sigma$ is obtained by Eq.\eqref{eq.sigma}. 
Using Eq.\eqref{equ7}, the new single-particle potentials $U_S^{(1)}$ and $U_0^{(1)}$ are obtained for the next iteration.

The binding energy per nucleon is
\begin{eqnarray}
    E/A &&  =  E_{T}/A  +  E_V/A,
\end{eqnarray}
where,
\begin{eqnarray}
    E_{T}/A &&    =      \frac{1}{\rho}
                         \sum_{\lambda}\int_0^{k_F}
                         \frac{\mathrm{d}^3p}{(2\pi)^3}
                         \frac{M^\ast}{E_{\bm{p}}^\ast}\langle\bar{u}(\bm{p},\lambda)|\bm{\gamma}\cdot\bm{p}+M|u(\bm{p},\lambda)\rangle-M,\nonumber\\
    E_V/A   &&    =      \frac{1}{2\rho}
                         \sum_{\lambda,\lambda'}
                         \int_0^{k_F}\frac{\mathrm{d}^3p}{(2\pi)^3}\int_0^{k_F}
                         \frac{\mathrm{d}^3p'}{(2\pi)^3}\frac{M^\ast}{E_{\bm{p}}^\ast}\frac{M^\ast}{E_{\bm{p}'}^\ast}\nonumber\\
            && \times    \langle \bar{u}(\bm{p},\lambda)\bar{u}(\bm{p}',\lambda')|\bar{G}(W)|u(\bm{p},\lambda)u(\bm{p}',\lambda')\rangle,\label{equ16}
\end{eqnarray}
 where $k_F$ is related to the total baryonic density $\rho = \rho_n  +  \rho_p$.

\section{\label{sec:level3} Results and discussions}
\subsection{$NN$ Phase shifts}

In the present work, the pion
and nucleon masses are fixed at $m_\pi =138.03$ MeV and
$M = 938.926$ MeV. 
The pion decay and the axial coupling constants of the nucleon are respectively $f_\pi=92.4$ MeV and $g_A=1.29$ \cite{Machleidt:2011zz}.

For the momentum cutoff $\Lambda$ from 450 to 600 MeV, the LECs corresponding to the short-range operators 
in Table \ref{Table1} are obtained by fitting  the PWA93 $np$ phase shifts \cite{Stoks:1993tb}.
The partial waves data $J\leq2$ with the laboratory kinetic energy $E_{\rm lab}=1,5,10,25,50,100,150,200$ MeV are considered.

Following the fitting procedure from the non-relativistic $\chi$EFT~\cite{Ordonez:1995rz}, all LECs are obtained by fitting the $np$ phase shifts by the $\chi^2$-like function
\begin{equation}    
  \tilde{\chi}^2  =  \sum_i(\delta^i-\delta_{\rm PWA93}^i)^2,\label{equ13}
\end{equation}
where $\delta^i$ are calculated phase shifts or mixing angles, and
$\delta_{\rm PWA93}^i$ are the PWA93 data.
The Migrad algorithm in the Minuit package are used to minimize the chi-square function.
However, the minima is found to be dependent on the initial conditions of the minimization.
Moreover, the resulting LECs strongly deviate from their expected naturalness \cite{Epelbaum:1998na,Epelbaum:2014efa}, i.e., the LECs satisfy
$\tilde{C}\sim1/{f}^{\hspace{0.1em}2}_{\pi}$.
This leads to unnaturally large interaction matrix elements and instabilities in the RBHF calculations for nuclear matter.

The determination of LECs in the covariant chiral nuclear force is not trivial.
There are two reasons.
Firstly, up to NLO, there are 17 LECs in the relativistic case, and only 9 LECs in the non-relativistic case.
The difference arises because the non-relativistic contact terms only contain momenta $\bm p$ and $\bm p'$ to the second power, 
while the relativistic contact terms also include higher powers of momenta dictated by Lorentz covariance.
The non-relativistic contact terms are listed in Table \ref{Table3}. 
Secondly, the fitting must be done simultaneously for all partial waves in the relativistic case, 
while it can be done separately for each partial wave in the non-relativistic case \cite{EPELBAUM2005362}.
Therefore, the LECs in the covariant chiral nuclear force form a high-dimensional parameter space, leading to numerical difficulties 
in the scattering phase shift fitting.
In fact, a similar problem in the non-relativistic chiral nuclear force arises 
at the next-to-next-to-next-to-leading order (N$^3$LO) when the number of LECs increases.
To solve this problem, the naturalness has been adopted, and additional physical constraints are introduced 
in the fitting procedure \cite{Epelbaum:2014efa}.

\begin{table}[!htbp]
\caption{Operators entering the LO ($Q^0$) and NLO ($Q^2$) contact interactions in the non-relativistic heavy-baryon chiral EFT \cite{Epelbaum:2004fk}.}\label{Table3}
\centering
  \begin{tabular}{cccc}
\hline\hline
$O_{\mathrm{S}}$       &    $\qquad$     &                                                 $1$                                                       \\
$O_{\mathrm{T}}$       &    $\qquad$     &                                 $\bm{\sigma}_1\cdot\bm{\sigma}_2$                                         \\
$O_1$                  &    $\qquad$     &                                         $p^2+p^{\prime2}$                                                 \\
$O_2$                  &    $\qquad$     &                                        $\bm{p}\cdot\bm{p}'$                                                \\
$O_3$                  &    $\qquad$     &                    $i(\bm{\sigma}_1+\bm{\sigma}_2)\cdot(\bm{p}\times\bm{p}')$                              \\
$O_4$                  &    $\qquad$     &                        $(p^2+p^{\prime2})\bm{\sigma}_1\cdot\bm{\sigma}_2$                                  \\
$O_5$                  &    $\qquad$     &                       $\bm{\sigma}_1\cdot\bm{\sigma}_2 \bm{p}\cdot\bm{p}'$                                 \\
$O_6$                  &    $\qquad$     &  $\bm{\sigma}_1\cdot\bm{p}\bm{\sigma}_2\cdot\bm{p}+\bm{\sigma}_1\cdot\bm{p}'\bm{\sigma}_2\cdot\bm{p}'$     \\
$O_7$                  &    $\qquad$     &  $\bm{\sigma}_1\cdot\bm{p}\bm{\sigma}_2\cdot\bm{p}'+\bm{\sigma}_1\cdot\bm{p}'\bm{\sigma}_2\cdot\bm{p}$     \\
\hline\hline
\end{tabular}
\end{table}

To overcome the difficulties for phase shift fitting in high-dimensional parameter space, 
the fitting strategy for the LECs in the covariant chiral nuclear forces include the following three-steps.
\begin{itemize}
\item  The four LO LECs ($\tilde{C}_{1\text{-}4}$) are fitted by the $np$ phase shift data for partial waves with $J\leq 1$. 
\item  Five NLO LECs, $\tilde{C}_{6}$, $\tilde{C}_{10}$, $\tilde{C}_{13}$, $\tilde{C}_{14}$, and $\tilde{C}_{16}$, are inculded into the fitting.  
       The nine LECs are fitted by the $np$ phase shift data for partial waves with $J\leq 2$.
       The rationale for selecting these nine LECs is that these nine contact terms recover the nine non-relativistic counterpart after non-relativistic reduction.  
       Although there are different choices for the five NLO LECs, the present choice yields the lowest $\tilde{\chi}^2$. 
\item  All 17 LECs up to NLO are fitted by the $np$ phase shift data for $J\leq2$ partial waves 
       with the 9 LECs initial values provided by the previous step. 
       The fittings are performed with the $\chi^2$-like function in Eq.\eqref{equ13} (denoted as LEC-I), 
       or the $\chi^2$-like function with constraint term $\chi^2_{\mathrm{prior}}$ in Eq.\eqref{equ15} (denoted as LEC-II).
\end{itemize}

\begin{figure}[h]
\includegraphics[width=0.45\textwidth]{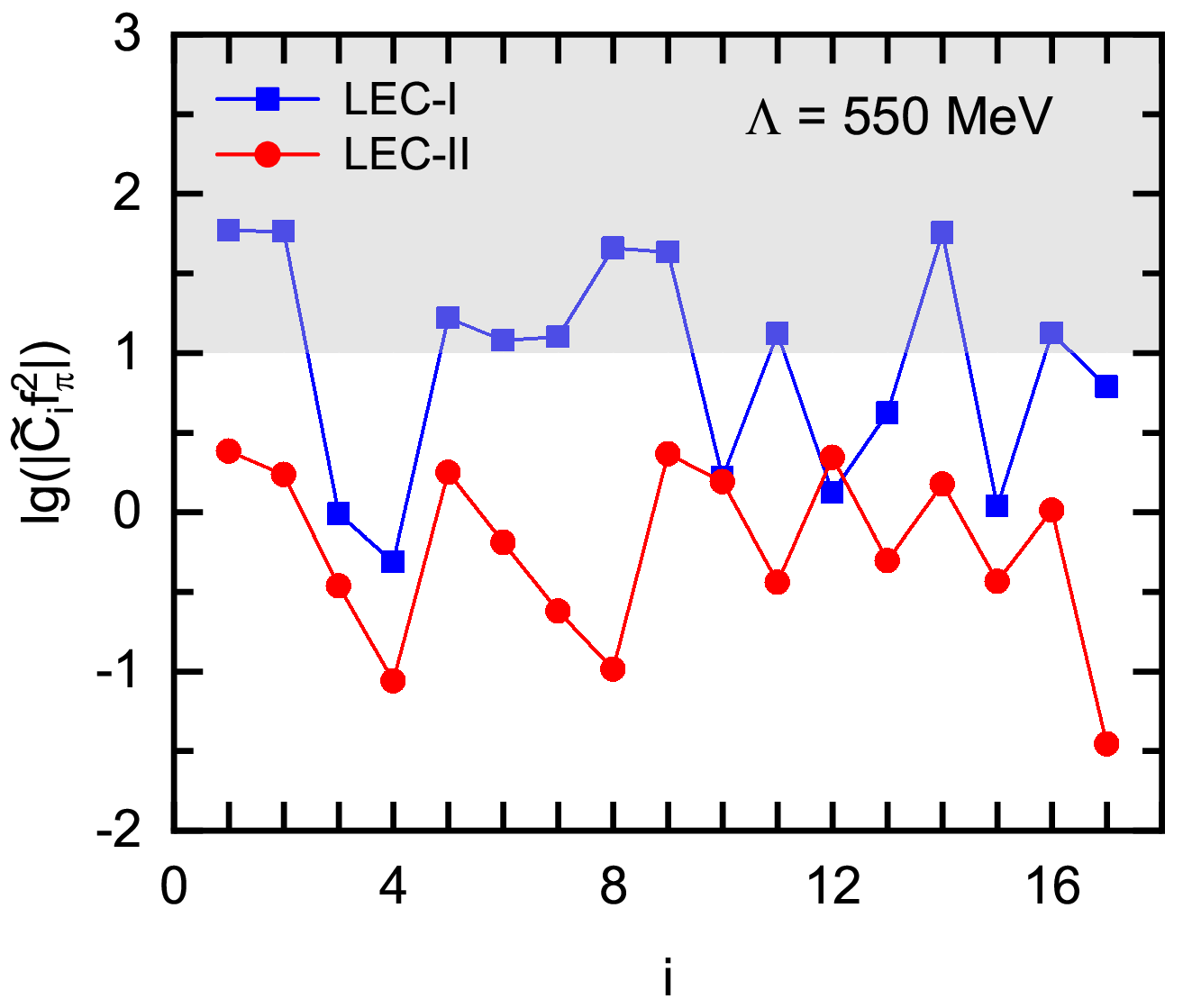}
\caption{(Color online) The relative magnitude of LECs with respect to the natural size $1/{f}^{\hspace{0.1em}2}_{\pi}$, 
obtained at the momentum cutoff $\Lambda=550$ MeV.
The LECs are obtained by different fitting strategies, denoted as LEC-I and LEC-II; see the text for details.
The shaded area indicates that the relative magnitudes of LECs exceed 10 when scaled against $1/{f}^{\hspace{0.1em}2}_{\pi}$.
}\label{fig1}
\end{figure}

For LEC-I, some of the obtained LECs are found to be large and significantly deviate from their natural size, as shown in fig.\ref{fig1}.

For LEC-II, in order to ensure the naturalness, we rewrite the contact terms as
\begin{align}
     V_{\rm CT}  = \sum_{i}(\tilde{C}_i + \sum_j \tilde{C}_jd_{ij})\tilde{O}_i + \sum_j\tilde{C}_j(\tilde{O}_j- \sum_i d_{ij}\tilde{O}_i).
\end{align}
Here, the 9 LECs in the previous steps are denoted by index $i=\{1,2,3,4,6,10,13,14,16\}$, the remaining 8 LECs are denoted by index $j=\{5,7,8,9,11,$
$12,15,17\}$, and the constant matrix $d_{ij}$ is given in Appendix~\ref{app}.  
Follow the previous steps, as the 9 LECs obtained satisfy the naturalness, we add the following constraint term $\chi^2_{\mathrm{prior}}$ in the $\chi^2$-like function
\begin{align}        
\chi^2_{\mathrm{prior}}  = \sum_{i}\frac{(\sum_j \tilde{C}_jd_{ij})^2}{(\zeta/2)^2},\label{equ15}
\end{align}
so that the 17 LECs satisfy the naturalness requirement. Here, 
$\zeta = 1/{f}^{\hspace{0.1em}2}_{\pi}$ is natural scale.

Taking the momentum cutoff $\Lambda=550$ MeV as an example, the relative magnitude $|\tilde{C}{f}^{\hspace{0.1em}2}_{\pi}|$ 
for the LECs in the LEC-I and LEC-II are given in Fig.\ref{fig1}.
In LEC-I, the relative magnitudes of LECs scaled by $1/{f}^{\hspace{0.1em}2}_{\pi}$ exceed 10, whereas in LEC-II, all LECs exhibit relative magnitudes below 10 when measured against the same scale.

\begin{figure}[h]
\includegraphics[width=0.45\textwidth]{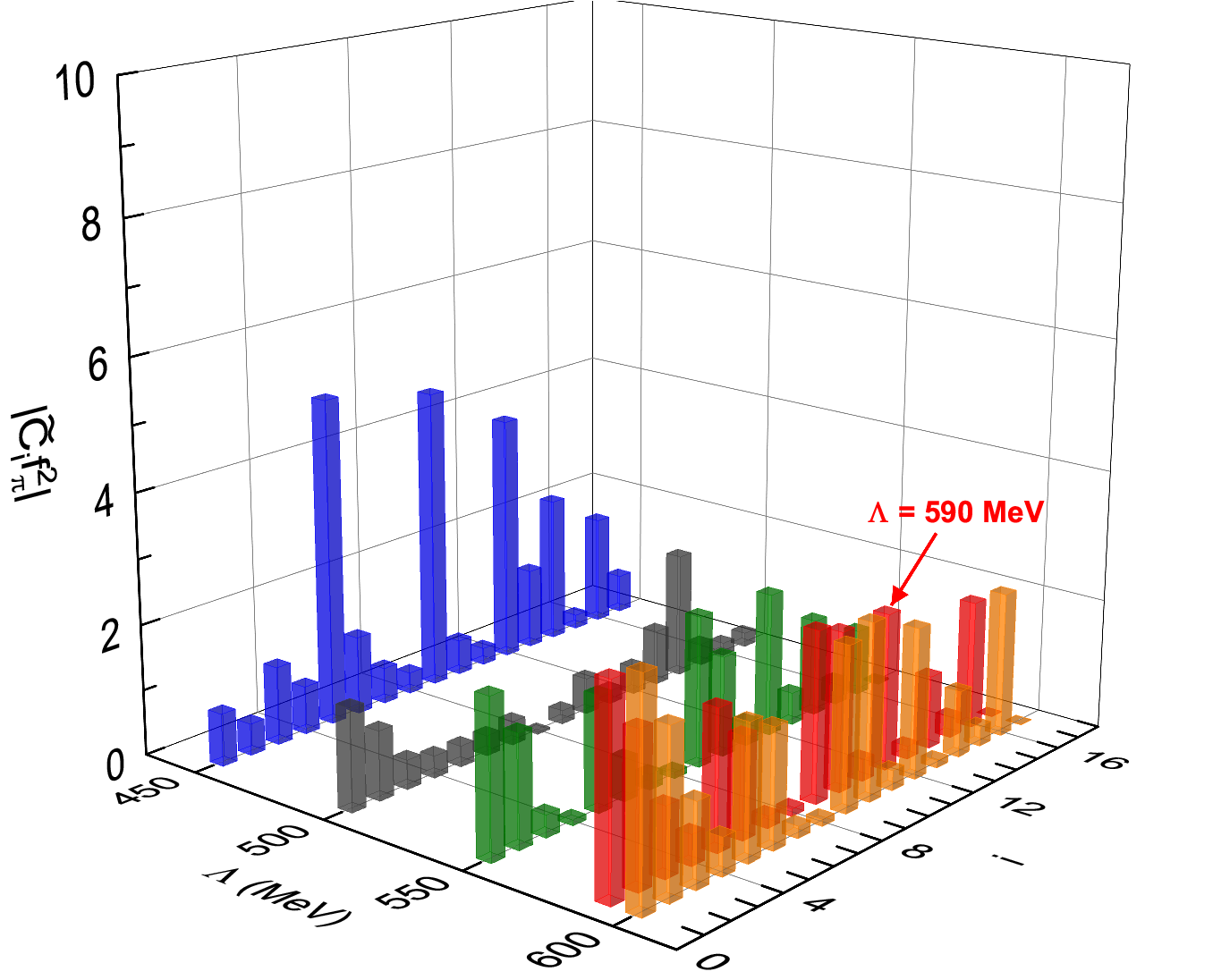}
\caption{(Color online) The relative magnitude of LECs with respect to the natural scale $1/{f}^{\hspace{0.1em}2}_{\pi}$ in the range of momentum cutoff $\Lambda=450\text{-}600$ MeV.}\label{fig2}
\end{figure}

In Fig.\ref{fig2}, we show the relative magnitude of the LECs obtained by LEC-II under different momentum cutoffs.
As shown in Fig.2, within the momentum cutoff range of $\Lambda=450\text{-}600$ MeV, the LEC-II consistently yields LECs with relative magnitudes below 10 when scaled against $1/{f}^{\hspace{0.1em}2}_{\pi}$.
The specific values of LECs are specified for momentum cutoff $\Lambda=590$ MeV in Table~\ref{Table2}.

\begin{table*}[h]
\centering
\caption{The values of LECs~(in units of $\mathrm{GeV}^{-2}$)~for the NLO covariant chiral nuclear force at momentum cutoff $\Lambda=590$ MeV. }\label{Table2} 
  \begin{tabular}{cccccccccc}
\hline\hline
$\tilde{C}_1$    &  $\tilde{C}_2$     &  $\tilde{C}_3$    &  $\tilde{C}_4$    &  $\tilde{C}_5$    &  $\tilde{C}_6$    &  $\tilde{C}_7$    &  $\tilde{C}_8$    & $\tilde{C}_9$  \\
\hline
$-369.58$        &  $270.97$          &  $-124.07$        &  $-54.67$         &  $-247.07$        &  $180.07$         &  $-22.09$         &   $13.51$         &  $-303.42$     \\
\hline
$\tilde{C}_{10}$ &  $\tilde{C}_{11}$  &  $\tilde{C}_{12}$ &  $\tilde{C}_{13}$ &  $\tilde{C}_{14}$ &  $\tilde{C}_{15}$ &  $\tilde{C}_{16}$ &  $\tilde{C}_{17}$                   \\
\hline
$-282.77$        &  $38.84$           &  $-278.53$        &  $-23.30$         &   $130.56$        &   $-37.92$        &   $-229.02$       &   $-0.18$         &   $\qquad$       \\       
\hline\hline
\end{tabular}
\end{table*}

\begin{figure*}[h]
\includegraphics[scale=0.70]{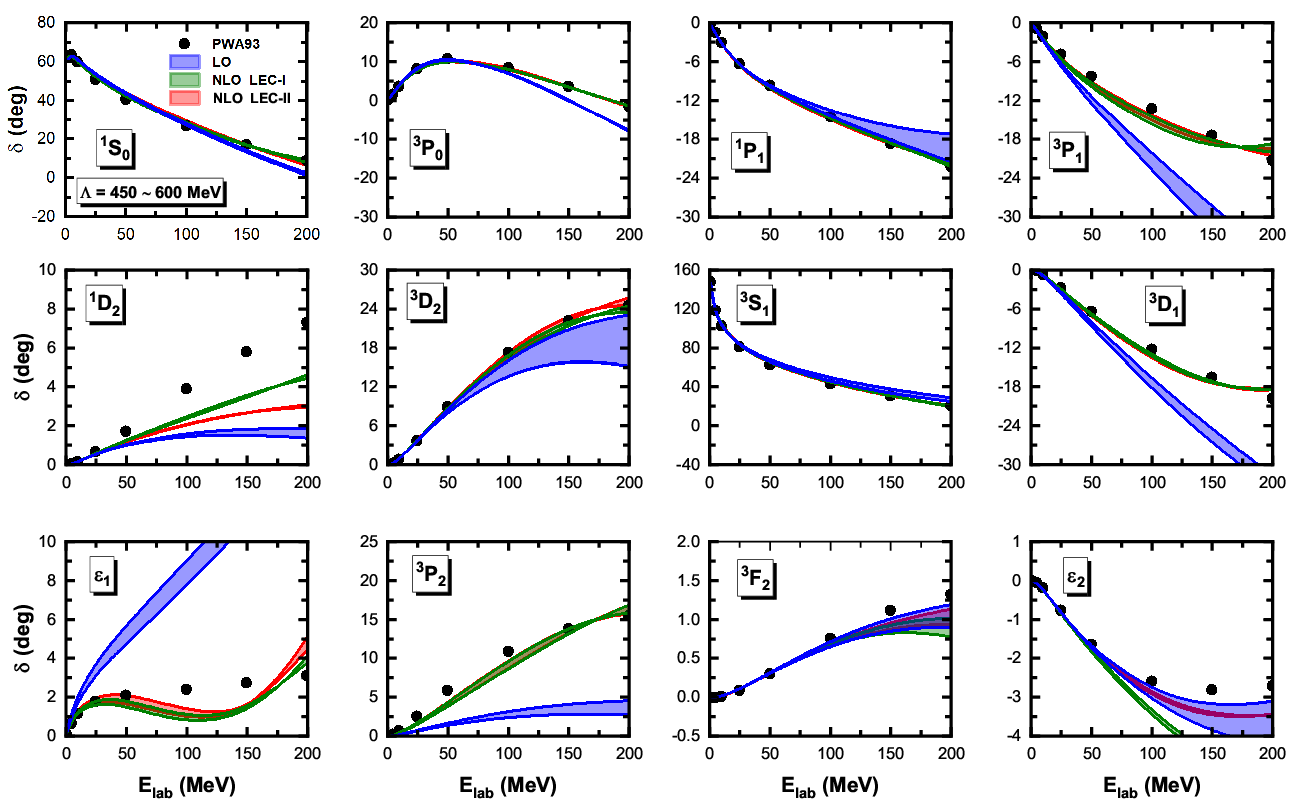}
\caption{(Color online) Neutron-proton phase shifts for partial waves with $J\leq2$. 
The LECs in the NLO covariant chiral nuclear force are obtained by different fitting strategies, 
denoted as LEC-I and LEC-II, see the text for details. The bands are obtained by varying the momentum cutoff $\Lambda$ from 450 to 600 MeV.}\label{fig3}
\end{figure*}
In Fig.\ref{fig3}, the descriptions of scattering phase shifts for partial waves with $J\leq2$ are compared among two different fitting schemes in the NLO, namely LEC-I and LEC-II. 
In addition, we also compare with the LO results. 
The results are shown for different momentum cutoffs $\Lambda$ from 450 to 600 MeV.  
The overall descriptions of the scattering phase shifts for partial waves with $J\leq2$ by the LEC-I and LEC-II are quite similar.
When moving from LO to NLO, the momentum cutoff dependence is reduced. 
Especially, the description of higher partial waves, such as the P-waves and D-waves is significantly improved at the NLO.

\subsection{Symmetric nuclear Matter}

\begin{figure}[h]
\includegraphics[width=0.45\textwidth]{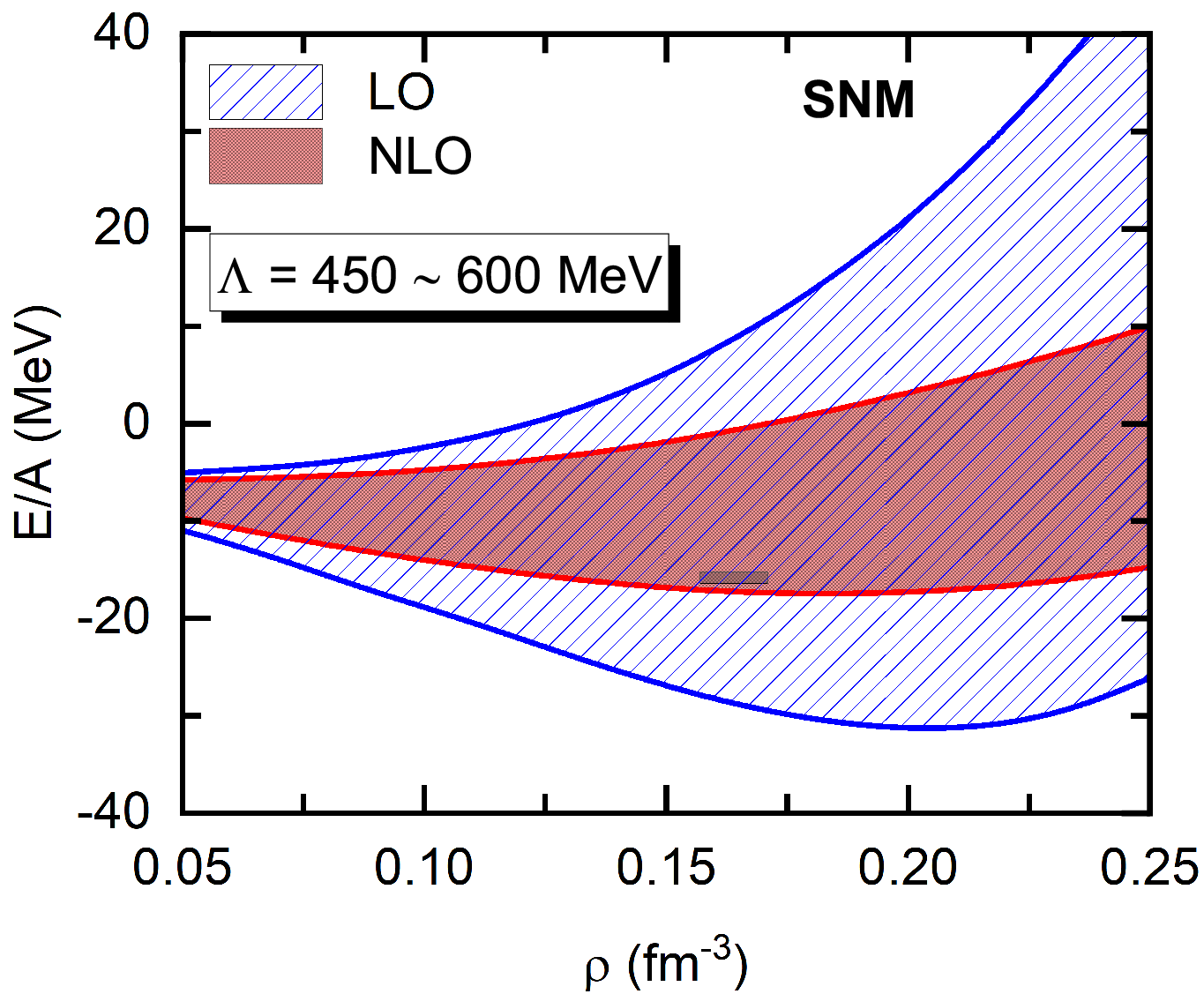}
\caption{(Color online)~Energy per nucleon~($E/A$)~in SNM as a function of the density~$\rho$~in the RBHF theory obtained with the covariant chiral nuclear force. 
The red band represents the NLO results, while the blue band represents the LO results taken from Ref.~\cite{Zou:2023quo}.
The bandwidth denotes the momentum cutoff variance in the range $\Lambda=450\text{-}600$ MeV.
}\label{fig4}
\end{figure}
In Fig.~\ref{fig4}, we show the momentum cutoff dependence of the EoSs for SNM obtained by the LO and NLO covariant chiral nuclear forces. 
The LECs in the LEC-I do not satisfy the naturalness requirements, and a convergent result in the RBHF calculations for nuclear matter cannot be obtained.
Therefore, from here and after, we present the results based on the LECs in the LEC-II.
The blue and red shaded bands represent our LO and NLO calculations, respectively. 
In all the cases shown, the momentum cutoff is varied over the range 450-600 MeV. 
The LO band completely contains the NLO band as expected, indicating that the NLO results exhibit a reduced momentum cutoff uncertainty.
Compared with the LO results, the EoSs obtained at the NLO become softer at densities $\rho>\rho_0$. 
This is consistent with the results for phase shifts, since the over-repulsion in the P-waves is improved at the NLO. 
The momentum cutoff uncertainty in the NLO calculation of the EoSs is reduced by half compared with the LO uncertainty. 
Although the NLO calculation reveals a strong reduction of the momentum cutoff uncertainty, there is still an uncertainty of $\sim$16 MeV at the saturation density.
This establishes that \textit{ab initio} calculations of nuclear matter using covariant chiral nuclear forces display systematic order-by-order convergence, 
while suggesting that the residual strong momentum cutoff dependence could be further mitigated at higher chiral orders.

\begin{figure}
\includegraphics[width=0.45\textwidth]{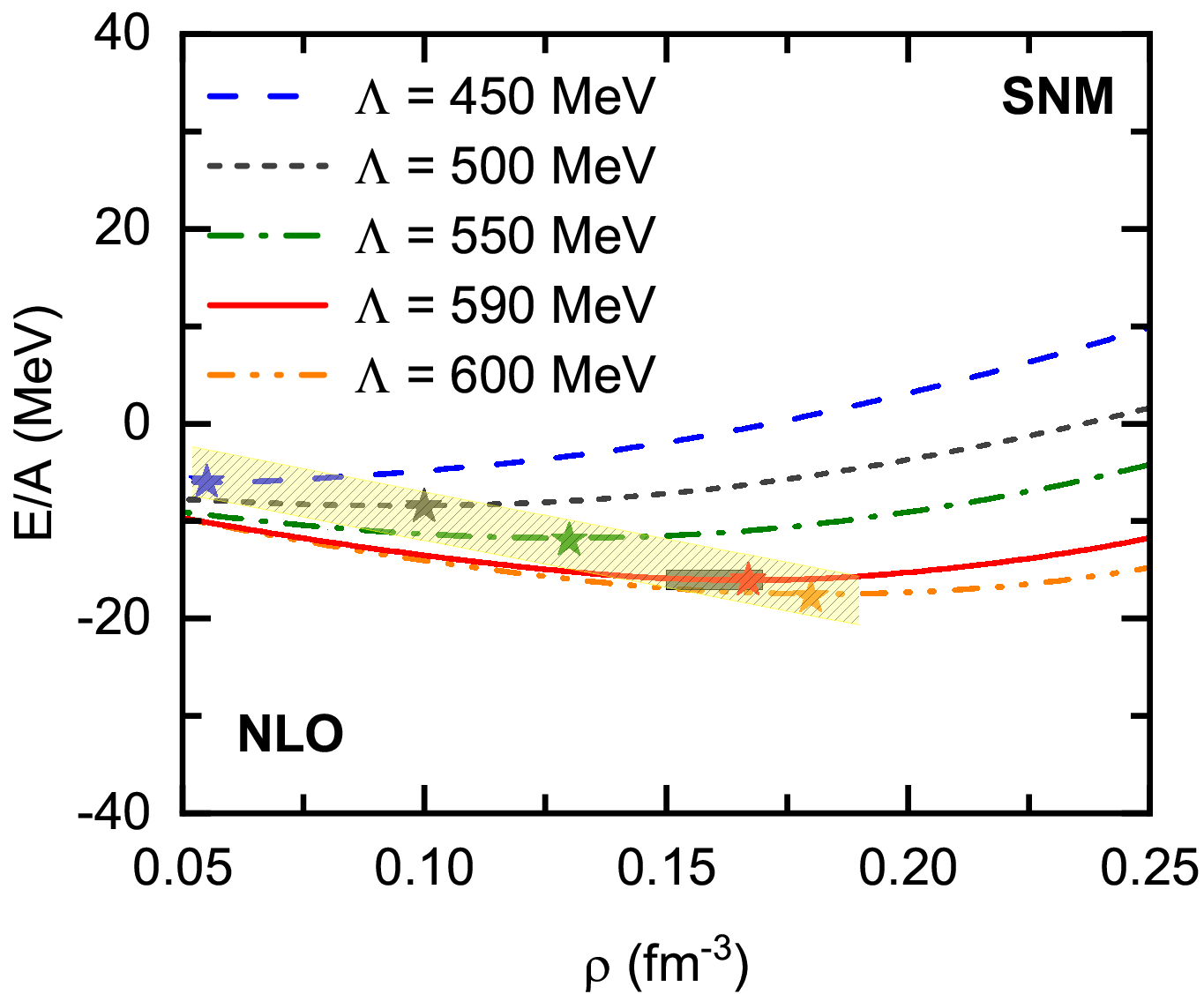}
\caption{(Color online) Energy per nucleon ($E/A$) in SNM as a function of the density $\rho$ in the RBHF theory obtained with the NLO covariant chiral nuclear force \cite{Lu:2021gsb}. 
The pentagrams denote the saturation point. The shaded area indicates the empirical value \cite{Bethe:1971xm}. 
}\label{fig6}
\end{figure}
In Fig.\ref{fig6}, we display the EoSs of SNM calculated by the NLO covariant chiral nuclear forces under different momentum cutoffs. 
The EoSs become increasingly repulsive as the momentum cutoff $\Lambda$ decreases. 
The saturation points calculated at different momentum cutoffs $\Lambda$ form a new “Coester line”, which lies closer to the empirical value.
Specifically, for $\Lambda=590$MeV, the energy per nucleon at saturation density obtained through the RBHF theory is $-16.05$ MeV, 
which is in excellent agreement with the empirical value of $-16\pm1$ MeV \cite{Bethe:1971xm}. 
Moreover, the calculated saturation density $\rho_0=0.167$ fm${}^{-3}$ is also in good agreement with the empirical value of $0.16\pm0.01$ fm${}^{-3}$ \cite{Bethe:1971xm}.

\begin{figure}[h]
\includegraphics[width=0.45\textwidth]{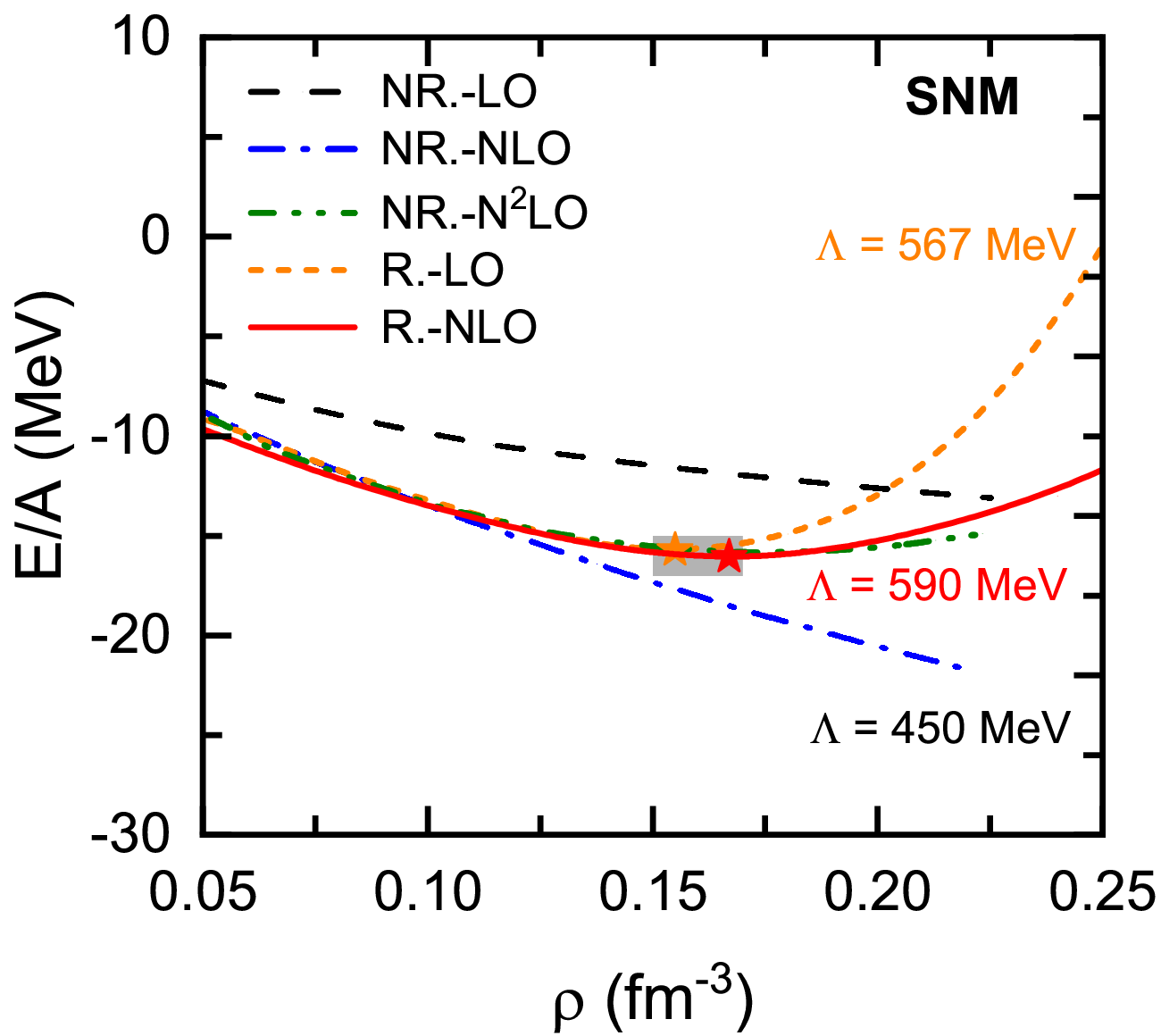}
\caption{(Color online)~Energy per nucleon ($E/A$) in SNM as a function of the density $\rho$ with the LO and NLO covariant chiral nuclear forces (yellow short dash line and red solid line), 
in comparison with the non-relativistic LO, NLO, and N${}^2$LO chiral nuclear forces (black dashed line, blue dotted dash line and green dot-dot-dashed line) \cite{Sammarruca:2021bpn}. 
The shaded area indicates the empirical value \cite{Bethe:1971xm}.
}\label{fig7}
\end{figure}
In Fig.\ref{fig7}, we compare the RBHF results obtained with the covariant chiral nuclear forces to the BHF results calculated 
using the non-relativistic chiral nuclear forces \cite{Sammarruca:2021bpn}. 
For the BHF results, there is no sign of saturation up to NLO in the density range considered here and only when 3NF is considered at the N${}^2$LO, 
a reasonable description of nuclear saturation can be achieved. 
In contrast, the RBHF results employing the NLO covariant chiral nuclear force 
with a momentum cutoff $\Lambda = 590$ MeV can describe the nuclear saturation reasonably well. 
For $\Lambda=590$ MeV, the incompressibility coefficient $K_\infty$ at the saturation density is $270$ MeV, 
which agrees fairly with the empirical value of $240\pm20$ MeV \cite{Garg:2018uam}. 
Compared with the LO results, the EoSs of SNM from the present NLO calculations are softer at densities above the saturation density.

\subsection{Pure neutron Matter}

\begin{figure}[h]
\includegraphics[width=0.45\textwidth]{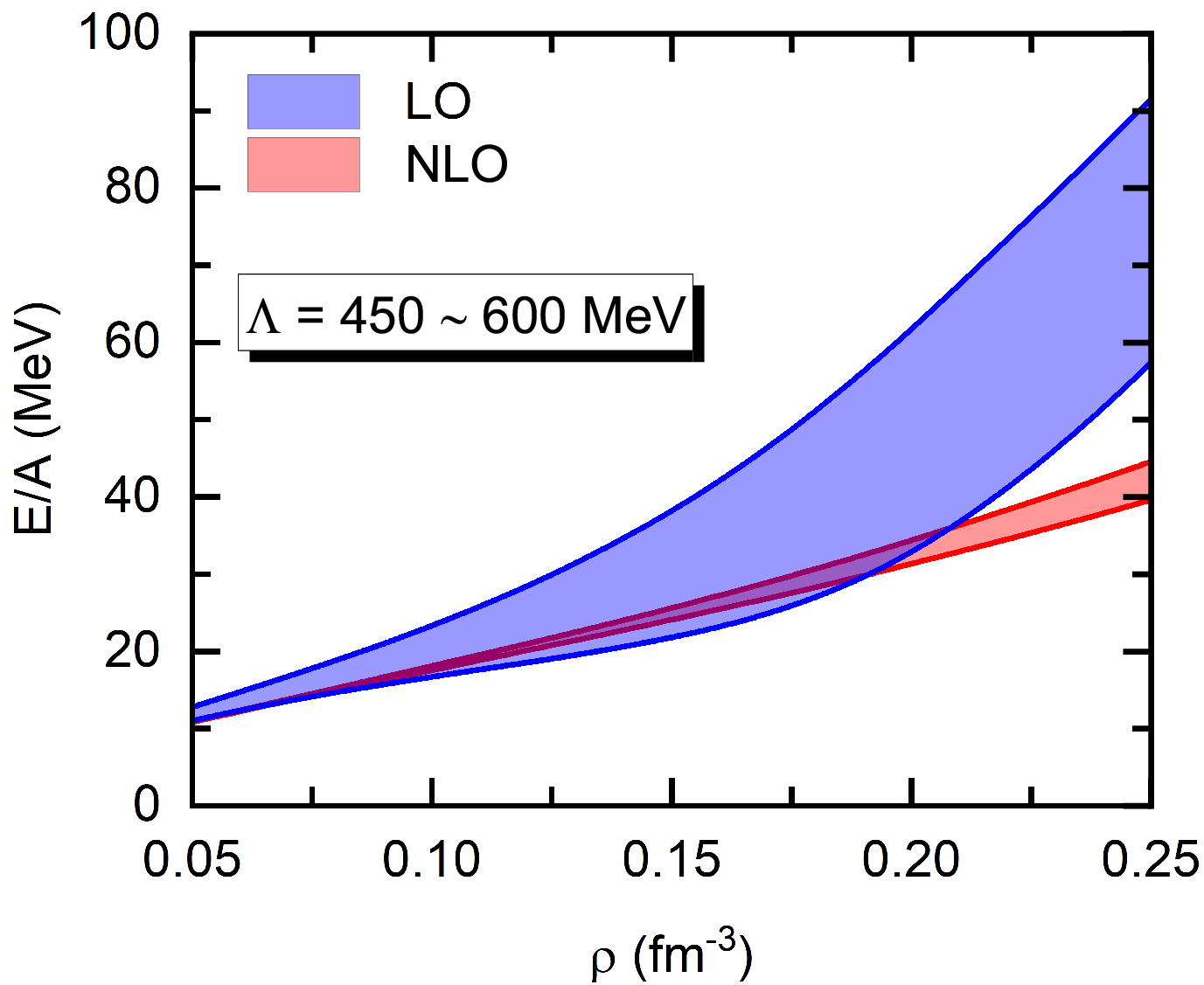}
\caption{(Color online) Energy per nucleon ($E/A$) in PNM as a function of the density $\rho$ calculated in the RBHF theory with the LO and NLO covariant chiral nuclear forces. 
The blue and red bands represent the uncertainties in the predictions due to cutoff variations as obtained in calculations at the LO and NLO, respectively. }\label{fig8}
\end{figure}
In Fig.\ref{fig8}, we show the momentum cutoff dependence of the EoSs for PNM obtained by the RBHF theory 
with the LO and NLO covariant chiral nuclear forces. 
The shaded bands in blue and red represent the EoSs for momentum cutoff $\Lambda=450\text{-}600$ MeV at the LO and NLO, respectively. 
At the LO, the momentum cutoff uncertainty increases rapidly as the densities become higher, and reaches about $\sim30\%$ at $\rho=0.25$ fm$^{-3}$.
At the NLO, the momentum cutoff uncertainty is considerably reduced compared to the LO, about $\sim10\%$ at $\rho=0.25$ fm$^{-3}$.
This is similar to what was observed in SNM.
The NLO and LO bands overlap with each other at $\rho<0.20$ fm$^{-3}$, while the LO EoSs are much more repulsive at higher densities. 
Compared with the LO results, the EoSs obtained at the NLO become significantly softer.
This is consistent with the results for phase shifts, since the over-repulsion in the P-waves at the LO is improved at the NLO. 

\section{\label{sec:level4} Summary and Outlook}

In summary, we present the RBHF calculations of the EoSs for SNM and PNM employing covariant chiral nuclear forces up to NLO.
For the NLO covariant chiral nuclear forces, the LECs in the contact interactions are determined by fitting to the $np$ scattering phase shifts 
with the Thompson equation, consistent with the scattering equation employed in the RBHF theory.
Directly fitting phase shifts with all 17 LECs up to NLO produced unnaturally large values, causing numerical instabilities in nuclear matter RBHF calculations.
Therefore, a fitting scheme to ensure the naturalness of LECs is proposed, while keeping the accuracy of describing the np phase shifts.

The NLO covariant chiral nuclear forces with the obtained LECs are then used in the RBHF theory to study the SNM and PNM.
For the SNM, the RBHF calculations with the NLO covariant chiral nuclear forces at momentum cutoff $\Lambda=590$ MeV yield saturation energy per nucleon $-16.05$ MeV and 
saturation density $0.167$ fm${}^{-3}$, consistent with the empirical values $-16\pm1$ MeV and $0.16\pm0.01$ fm${}^{-3}$.
In addition, the calculated incompressibility coefficient $K_\infty$ at the saturation density is $270$ MeV, 
which agrees fairly with the empirical value of $240 \pm 20$ MeV. 
Given the good description for the saturation properties of nuclear matter, the present work encourages future studies of 
the finite nuclei in the framework of the RBHF theory with the NLO covariant chiral nuclear forces.

The momentum cutoff dependence of the EoSs is studied by varying the momentum cutoff scale in the range $\Lambda=450\text{-}600$ MeV.
The momentum cutoff uncertainty in the NLO calculation of the nuclear matter EoS is reduced by half compared with the LO uncertainty, 
as expected from the systematic order-by-order convergence of chiral expansion.
The saturation points obtained at different cutoffs lie in a band that overlaps with the empirical saturation region.
Nevertheless, the momentum cutoff dependence is still significant, and it should be interesting to extend 
the present study to higher orders to investigate how the higher-order contributions can reduce such momentum cutoff dependence.

For the PNM, the calculated EoSs at the LO and NLO overlap with each other below $\rho=0.2$ fm$^{-3}$.
At higher densities, the NLO EoSs become softer compared to the LO ones, due to the fact that the overestimation of the repulsion 
in the P-waves at the LO is significantly improved at the NLO.
The momentum cutoff uncertainty of the EoSs for PNM is much smaller than that in SNM.
In the future, it would be interesting to extend the present calculations of pure neutron matter to much higher densities 
to study the neutron star properties.

\section*{Acknowledgments}
This work was supported in part by the National Natural Science Foundation of China under Grants No.12435006,
No.12435007,
No.12475117, No.12141501, No.123B2080, the National Key R\&D Program of China under Grant No.2024YFA1612600, No.2024YFE0109803,
No.2023YFA1606703, and the National Key Laboratory of Neutron Science and Technology NST202401016.

\newpage

\appendix
\section{Appendix}\label{app}
\begin{widetext}
The expressions of the contact interactions in the NLO covariant chiral nuclear force after non-relativistic reduction are shown in Table~\ref{Table4}
\begin{table}[h]
\centering
\caption{The non-relativistic expressions corresponding to the contact interactions of Table~\ref{Table1}}\label{Table4}
  \begin{tabular}{cccc}
\hline\hline
$\tilde{O}_1$     &   $\qquad$     &     $ O_S + \frac{1}{4M^2}(-O_1-2O_2+2O_3) $                            \\
$\tilde{O}_2$     &   $\qquad$     &     $ O_S + \frac{1}{4M^2}(-2O_1+4O_2-6O_3+O_4+2O_5+O_6+2O_7)$           \\
$\tilde{O}_3$     &   $\qquad$     &     $-O_T + \frac{1}{4M^2}(-2O_3+O_4+2O_5-O_6+6O_7)$                     \\
$\tilde{O}_4$     &   $\qquad$     &     $2O_T + \frac{1}{4M^2}(2O_1+4O_2-12O_3-4O_4+8O_5-2O_6+12O_7)$        \\
$\tilde{O}_5$     &   $\qquad$     &     $     - \frac{1}{4M^2}(O_6+2O_7) $                                   \\
$\tilde{O}_6$     &   $\qquad$     &     $       \frac{1}{4M^2}(-4O_6+8O_7)$                                   \\
$\tilde{O}_7$     &   $\qquad$     &     $       \frac{1}{4M^2}(O_1+2O_2-8O_3-4O_4+8O_5-4O_6+8O_7)$            \\
$\tilde{O}_8$     &   $\qquad$     &     $       \frac{1}{4M^2}(-2O_1-4O_2+8O_3)$                             \\
$\tilde{O}_9$     &   $\qquad$     &     $       \frac{1}{4M^2}(O_4+2O_5+O_6+2O_7)$                           \\
$\tilde{O}_{10}$  &   $\qquad$     &     $       \frac{1}{4M^2}(O_1+2O_2)$                                    \\
$\tilde{O}_{11}$  &   $\qquad$     &     $       \frac{1}{4M^2}(O_1+2O_2)$                                    \\
$\tilde{O}_{12}$  &   $\qquad$     &     $       \frac{1}{4M^2}(-O_4-2O_5)$                                    \\
$\tilde{O}_{13}$  &   $\qquad$     &     $       \frac{1}{4M^2}(2O_4+4O_5)$                                     \\
$\tilde{O}_{14}$  &   $\qquad$     &     $       \frac{1}{4M^2}(-3O_1+2O_2)$                                    \\
$\tilde{O}_{15}$  &   $\qquad$     &     $       \frac{1}{4M^2}(-3O_1+2O_2)$                                    \\
$\tilde{O}_{16}$  &   $\qquad$     &     $       \frac{1}{4M^2}(3O_4-2O_5)$                                     \\
$\tilde{O}_{17}$  &   $\qquad$     &     $       \frac{1}{4M^2}(-6O_4+4O_5)$                                    \\
\hline\hline
\end{tabular}
\end{table}

The coefficient matrix representing the nine relativistic operator structures selected in the second step in terms of non-relativistic operator structures is
\begin{align}
\begin{pmatrix}
\tilde{O}_1\\
\tilde{O}_2\\
\tilde{O}_3\\
\tilde{O}_4\\
\tilde{O}_6\\
\tilde{O}_{10}\\
\tilde{O}_{13}\\
\tilde{O}_{14}\\
\tilde{O}_{16}\\
\end{pmatrix}=
\begin{pmatrix}
$1$  &   $0$  &   $-1$  &  $-2$  &  $2$   &  $0$  &  $0$  &  $0$  &  $0$ \\
$1$  &   $0$  &   $-2$  &  $4$   &  $-6$  &  $1$  &  $2$  &  $1$  &  $2$ \\
$0$  &   $-1$ &   $0$   &  $0$   &  $-2$  &  $1$  &  $2$  &  $-1$ &  $6$ \\
$0$  &   $2$  &   $2$   &  $4$   &  $-12$ &  $-4$ &  $8$  &  $-2$ &  $12$ \\
$0$  &   $0$  &   $0$   &  $0$   &  $0$   &  $0$  &  $0$  &  $-4$ &  $8$ \\
$0$  &   $0$  &   $1$   &  $2$   &  $0$   &  $0$  &  $0$  &  $0$  &  $0$ \\
$0$  &   $0$  &   $0$   &  $0$   &  $0$   &  $2$  &  $4$  &  $0$  &  $0$ \\
$0$  &   $0$  &   $-3$  &  $2$   &  $0$   &  $0$  &  $0$  &  $0$  &  $0$ \\
$0$  &   $0$  &   $0$   &  $0$   &  $0$   &  $3$  &  $-2$ &  $0$  &  $0$
\end{pmatrix}
\begin{pmatrix}
O_S\\
O_T\\
O_1\\
O_2\\
O_3\\
O_4\\
O_5\\
O_6\\
O_7
\end{pmatrix}
\end{align}
By inverting the above coefficient matrix, the result of representing the non-relativistic operator structures in terms of relativistic operator structures can be obtained.
\begin{align}
\begin{pmatrix}
O_S\\
O_T\\
O_1\\
O_2\\
O_3\\
O_4\\
O_5\\
O_6\\
O_7
\end{pmatrix}=
\begin{pmatrix}
\frac{1}{2}    &    \frac{1}{2}  &  -\frac{1}{4}  &  -\frac{1}{8}  &    \frac{1}{4}   &    \frac{1}{4}   &       0       &    -\frac{1}{2}    &   -\frac{1}{4}  \\
\frac{1}{2}    &   -\frac{1}{2}  &  -\frac{1}{4}  &   \frac{3}{8}  &   -\frac{1}{4}   &    \frac{1}{4}   &       0       &     \frac{1}{2}    &    \frac{3}{4}   \\
    0          &        0        &        0       &         0      &          0       &    \frac{1}{4}   &       0       &    -\frac{1}{4}    &          0       \\
    0          &        0        &        0       &         0      &          0       &    \frac{3}{8}   &       0       &     \frac{1}{8}    &          0        \\
\frac{1}{4}    &   -\frac{1}{4}  &   \frac{1}{8}  &   \frac{1}{16} &    -\frac{1}{8}  &    \frac{3}{8}   &       0       &     \frac{1}{4}    &      \frac{1}{8}  \\
    0          &        0        &        0       &         0      &          0       &         0        &  \frac{1}{8}  &           0        &      \frac{1}{4}  \\
    0          &        0        &        0       &         0      &          0       &         0        &  \frac{3}{16} &           0        &     -\frac{1}{8}  \\
\frac{1}{2}    &   -\frac{1}{2}  &  \frac{1}{2}   &    \frac{1}{4} &    -\frac{5}{8}  &    \frac{1}{2}   & -\frac{1}{4}  &     \frac{1}{2}    &      \frac{1}{2}  \\
\frac{1}{4}    &   -\frac{1}{4}  &  \frac{1}{4}   &    \frac{1}{8} &   -\frac{3}{16}  &    \frac{1}{4}   & -\frac{1}{8}  &     \frac{1}{4}    &      \frac{1}{4}
\end{pmatrix}
\begin{pmatrix}
\tilde{O}_1\\
\tilde{O}_2\\
\tilde{O}_3\\
\tilde{O}_4\\
\tilde{O}_6\\
\tilde{O}_{10}\\
\tilde{O}_{13}\\
\tilde{O}_{14}\\
\tilde{O}_{16}\\
\end{pmatrix}
\end{align}
Using the above coefficient matrix, we can represent the contributions of the
second power and below of $\bm{p}$ and $\bm{p}'$ in the remaining eight LECs in the third step. By subtracting their contributions, 
we can obtain the contributions of the fourth and higher powers of $\bm{p}$ and $\bm{p}'$ for the eight relativistic contact terms.

\end{widetext}

\bibliography{apssamp}

\end{document}